\documentclass[jap,preprint,preprintnumbers,amsmath,amssymb,nobibnotes,showkeys,aip]{revtex4}

\usepackage{graphicx}
\usepackage{dcolumn}
\usepackage{bm}

\begin{document}


\title{Charge Retention in Quantized Energy Levels of Nanocrystals}

\author{Aykutlu D\^{a}na}
\email{aykutlu@fen.bilkent.edu.tr}

\author{\.{I}mran Ak\c{c}a}

\author{Or\c{c}un Ergun}

\author{Atilla Aydinli}
 \affiliation{Department of
Physics, Bilkent University, 06800 Ankara, Turkey}

\author{Ra\c{s}it Turan}
\affiliation{Department of Physics, Middle East Technical
University, 06800 Ankara, Turkey}

\author{Terje G. Finstad}
\affiliation{University of Oslo, Department of Physics, P.Box 1048
- Blindern, 0316 Oslo, Norway}

\date{\today}

\begin{abstract}
Understanding charging mechanisms and charge retention dynamics of
nanocrystal memory devices is important in optimization of device
design. Capacitance spectroscopy on PECVD grown germanium
nanocrystals embedded in a silicon oxide matrix was performed.
Dynamic measurements of discharge dynamics are carried out. Charge
decay is modelled by assuming storage of carriers in the ground
states of nanocrystals and that the decay is dominated by direct
tunnelling. Discharge rates are calculated using the theoretical
model for different nanocrystal sizes and densities and are
compared with experimental data. Experimental results agree well
with the proposed model and suggest that charge is indeed stored
in the quantized energy levels of the nanocrystals.
\end{abstract}


\maketitle
\section{Introduction}
The observation of formation of nanocrystals (NCs) by annealing of
silicon dioxide films having excess Si or Ge has attracted
attention due to optical and electronic properties of such
nanostructures
~\cite{introductionref1,introductionref2,introductionref3,gencs3}.
In particular, Ge NCs embedded in amorphous silicon oxide (a-SiO)
films have been subject of study, because of low temperature of
formation, compatibility with standard integrated circuit
fabrication processes and for their potential applications in
optoelectronic and memory devices. The NCs are candidates as
storage media for electron storage cells in flash memory
devices~\cite{geflash,singleelectron}. Since many parameters of
NCs such as density, size and composition can be adjusted by
proper choice of fabrication parameters, they offer flexibility in
design of NC flash memory cells.  However, a better understanding
of charge storage mechanism is important in optimization of device
performance. Recent studies have proposed a model describing the
storage of carriers in NC-MOS devices assuming storage in
deep-traps \cite{choi,choietal,sheking} associated with NCs and
trap energy level engineering was investigated to improve device
performance.

In this paper, we investigate an alternative mechanism for carrier
storage by assuming carrier storage in NC energy levels instead of
deep traps. Based on this assumption, we present a theoretical
model that includes the effect of NC dimensions and density to
calculate the discharge dynamics. Germanium NC-MOS capacitors have
been fabricated and characterized using capacitance measurements.
Results are compared with theory, showing agreement on size and
density related discharge properties.

\section{Theoretical modelling}
\begin{figure}[h]
\begin{center}
\resizebox{8 cm}{!}{\includegraphics{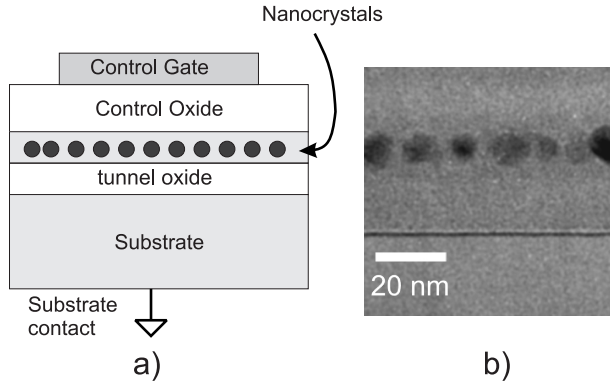}}
\end{center}
\caption{a) Schematic cross-section of a nanocrystal MOS capacitor
and b) example TEM micrograph of a calibration sample showing
germanium nanocrystal band with 7.4 nm average diameter
nanocrystals. } \label{MOSCAPandTEM}
\end{figure}

A typical NC memory element cross section is shown in Fig.
\ref{MOSCAPandTEM}. Based on the assumption that only NCs are
responsible for charge storage, the flat-band voltage shift
$\Delta V_{FB}$ is approximately given by
\cite{flatbandshiftreference}
\begin{equation}\label{hysteresisvoltage}
\Delta V_{FB}=
\frac{q_{nc}}{\epsilon_{ox}}(t_{cox}+\frac{\epsilon_{ox}t_{nc}}{2\epsilon_{ge}})
\end{equation}
where $q_{nc}$ is the total stored charge in the NCs, $t_{cox}$ is
the control oxide thickness, $t_{nc}$ is the average diameter of
the NCs, $\epsilon$'s are the dielectric constants of respective
materials. An important parameter of the NC-MOS device is the
maximum flat-band voltage shift $\Delta
V_{max}=qN_{NC}(t_{cox}+\epsilon_{ox}t_{nc}/{2\epsilon_{ge}})/\epsilon_{ox}$.
This is the flat-band voltage shift when all available NCs of
density $N_{nc}$ carry an electron i.e. $q_{nc}=qN_{nc}$. It is
seen that $\Delta V_{max}$ depends on device geometry through Eq.
\ref{hysteresisvoltage} and also on NC density. Due to large
Coulomb charging energy, average number of electrons per NC can be
assumed to be smaller than one.

In order to evaluate retention properties of NC-MOS memory
elements, discharging currents must be calculated. Since there are
many device parameters that collectively determine the
charge-discharge currents, a simple closed form formula can not be
obtained that covers all cases. Therefore retention and erase
currents are addressed separately.

\begin{figure}[h]
\begin{center}
\resizebox{7 cm}{!}{\includegraphics{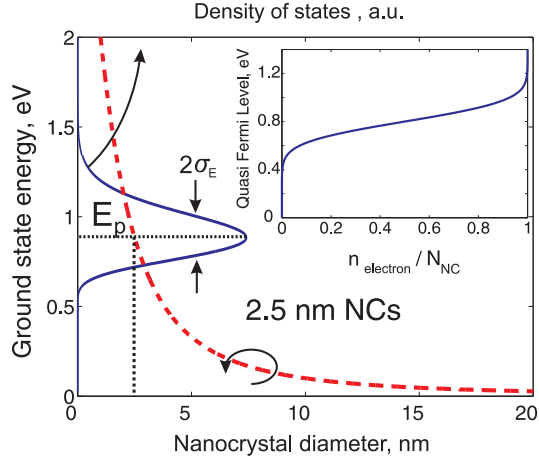}}
\end{center}
\caption{Schematic description of density of states (solid curve)
for the ground states of NCs with average diameter of 2.5 nm.
Dotted curve shows electron ground state of NCs as a function of
size as described by Eq. \ref{GENCCBM}. Inset shows the
quasi-Fermi level as a function of number of electrons per
nanocrystal.} \label{DOSNC}
\end{figure}

During retention, the device is in depletion and $V_{gate}=0$. If
NC bound states are responsible for storage of carriers, discharge
occurs by tunnelling from the NC ground state to the substrate,
either by direct or trap assisted tunnelling. For the calculation
of the discharge current, the barrier height of tunnelling
carriers must be calculated. The barrier height is a function of
the NC ground state energy given by $V_B(E)=V_{B0}-E_{nc}$, where
$V_{B0}$ is the bulk barrier height and $E_{nc}$ is the energy of
electron stored in the NC. The energy levels of uncapped germanium
NCs have recently been measured directly as a function of size,
using scanning tunnelling spectroscopy. The conduction band
minimum of Ge NCs as a function of size is given by
\cite{GeNCLevels}
\begin{eqnarray}\label{GENCCBM}
 E_{CBM}(d)=E_{CBM}(\infty)+ \frac{11.86}{d_{nc}^2+1.51d_{nc}+3.3936}
\end{eqnarray}
where the energies are in eV, $d_{nc}$ is the NC diameter in nm.
If we assume a Gaussian size distribution for the NCs, the density
of states, $D_{NC}(E)$, can then be calculated through Eq.
\ref{GENCCBM} for electrons as plotted in Fig. \ref{DOSNC}.

Assuming thermal equilibrium within the NC layer, the quasi-Fermi
level can be calculated implicitly (inset of Fig. \ref{DOSNC}) for
a given total stored charge. Escape of carriers near or above the
quasi-Fermi level dominates the discharge current. As a result of
reduced barrier height at the quasi-Fermi level for a large number
of carriers per NC, discharge current increases with the number of
stored carriers (or the flat-band voltage shift). This reduction
in barrier height, along with the increase in the tunnel oxide
field, results in the super-exponential charge decay commonly
observed in NC-MOS memory elements.

The current density describing the discharge of the NCs can be
calculated assuming direct tunnelling. For carriers stored in NCs
with density $n(E)$ all at an energy $E$, the discharge current
density $J_{d}$ can be given as
\begin{equation}\label{dischargecurrentdensity}
    J_{d}=qT_t(E,F_{tox})\nu_{NC}n(E)
\end{equation}
where $\nu_{NC}\simeq \hbar\pi/2m_{ge}d_{nc}^2$ is the
semi-classical escape attempt rate for NCs of diameter $d_{nc}$
\cite{attemptrate}. Here $T_t(E,F_{tox})$ is the barrier
transparency, for electrons with energy E and tunnel oxide field
$F_{tox}$. The actual discharge current must be obtained by
integration of Eq. \ref{dischargecurrentdensity} multiplied by the
density of states (DOS). The resulting current will be dominated
by tunnelling of carriers near the quasi-Fermi level. The
transmission probability $T_t(E,F_{tox})$ can be calculated
through the WKB approximation as
\cite{FNcurrent,oxidelowering,gehring}
\begin{equation}\label{transmissioncoefficient}
    T_t(E)\approx4\exp[-(1-(1-\frac{F_{tox}t_{tox}}{V_B(E)})^{3/2})\frac{BV_B(E)^{3/2}}{F_{tox}}].
\end{equation}
where $B=4\sqrt{2m_{ox}q}/3\hbar$, $m_{ox}$ being the electron
tunnelling mass. The tunnel oxide field $F_{tox}$ is determined by
the amount of stored carriers as well as by the band-bending . If
band-bending and gate-substrate work function difference is
ignored, the tunnel oxide field is approximately proportional to
the flat-band shift (or the total stored charge). Tunnel oxide
field is then given roughly by $F_{tox}\approx\Delta
V_{FB}/2t_{ox}$. Exact value of $F_{tox}$ depends on device
geometry and properties as well as band-bending.

During the erase cycle, $V_{gate}$ is negative and the device is
in inversion. The discharge current is determined by Eqs.
\ref{dischargecurrentdensity} and \ref{transmissioncoefficient}.
However, the oxide field is determined by the applied gate voltage
and flat-band voltage shift and is given approximately by
$F_{tox}\approx (-V_{gate}+\Delta V_{FB}/2)/t_{ox}$. A more
accurate description of $F_{tox}$ as a function of gate bias can
be obtained in a numerical calculation by taking into account the
band-bending of the substrate. Using the discharge currents given
in Eq. \ref{dischargecurrentdensity} and a standard band-bending
model \cite{nicollian}, the charge and discharge currents can been
calculated numerically.

\section{Experimental}
The oxide-germanosilicate-oxide trilayer films were grown in a
PECVD reactor (model PlasmaLab 8510C) on Si substrates using 180
sccm SiH$_{4}$ (2\% in N$_{2}$), 225 sccm NO$_{2}$ and varying
flow rates of GeH$_{4}$ (2\% in He) as precursor gases, at a
sample temperature of 350 $^{o}$C, a process pressure of 1000
mTorr under and an applied RF power of 10 W.  The samples were
then annealed in N$_{2}$ atmosphere in an alumina oven at
temperatures ranging from 650 $^o$C to 950 $^o$C for 5 minutes.
The samples were loaded and unloaded with ramp times of 1 minute.
For fabrication of the devices, first a thermal tunnel oxide of
thickness 4 nm was grown using dry oxidation on n-type silicon
substrates with resistivity of 1-10 $\Omega$cm, followed by PECVD
growth of germanosilicate layer of 10 nm thickness and composition
of $\rm{Si_{0.6}Ge_{0.4}O_2}$.  On top, a $\rm{t_{cox}=17}$ nm
control oxide was deposited. After annealing, backside ohmic
metallization and gate metallization was done by metal
evaporation.

Transmission electron microscopy (TEM) was used to characterize
the formation of NCs as a function of annealing temperature. High
density NC formation is observed for layers with a composition of
$\rm{Si_{0.6}Ge_{0.4}O_2}$ as determined by XPS analysis. The NC
diameter increases nonlinearly from 2.5 nm to 7.4 nm as the
annealing temperature is increased from 650 $^{o}$C to 850 $^{o}$C
as tabulated in Table \ref{nctable} for four devices. The average
energy, energy distribution width and maximum flat-band voltage
shift calculated using Eq.\ref{GENCCBM} and
Eq.\ref{hysteresisvoltage} are given in Table \ref{ncenergytable}
for the same devices in Table \ref{nctable}.

\begin{table}
\caption{Average nanocrystal size and width of size distribution
for different annealing temperatures as observed by TEM. \\}

\label{nctable}
\begin{tabular}{|c|c|c|c|c|}
  \hline
  Annealing & Average diameter  & size width&$N_{nc}$ density \\
  temperature ($^oC$) & (nm)&2$\sigma$ (nm)&($cm^{-2}$)\\
  \hline
  650 & 2.5 & 0.6&$8\times10^{12}$ \\
  700 & 2.8 & 0.7&$5.3\times10^{12}$ \\
  770 & 3.2 & 1.0 &$3.2\times10^{12}$\\
  850 & 7.4 & 1.6 & $8\times10^{11}$\\
  \hline
\end{tabular}
\end{table}

\begin{table}
\caption{Calculated properties of NCs based on size distribution
data.\\ } \label{ncenergytable}
\begin{tabular}{|c|c|c|c|c|}
  \hline
  Annealing & $E_p (eV)$& 2$\sigma_E$ (eV)&$\Delta V_{max}$\\
  temperature ($^oC$)& peak energy&
  width&(V)\\
  \hline
  650 & 0.88&0.55&16.9\\
  700 & 0.78&0.52&11.2\\
  770 & 0.64&0.53&6.3\\
  850 & 0.17&0.17&1.7\\
  \hline
\end{tabular}
\end{table}

Capacitance measurements were performed using a capacitance meter
(HP 4278A) with  1 MHz AC excitation of 25 mV amplitude. The
flat-band voltage shift can be tracked quasi-real-time for small
changes in the flat-band shift by using a digital feedback loop
that eliminates the need of tracing the whole C-V curve to
estimate the value of the flat-band voltage shift. During
write/erase pulses, the loop can be momentarily turned off. This
method allows rapid monitoring of the changes in the flat-band
voltage shift (within few tens of msec) after a write or erase
pulse or during retention.
\section{Results and Discussion}

Dynamic C-V measurements have been performed on NC-MOS capacitors,
by measuring the C-V as a function of time near the flat-band
voltage between applied pulses of varying voltage and durations.

\begin{figure}[h]
\begin{center}
\resizebox{7
cm}{!}{\includegraphics{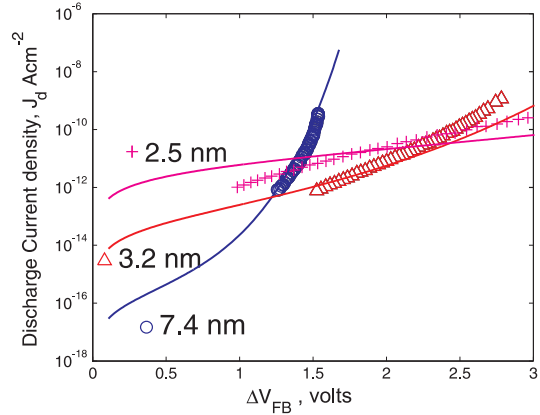}}
\end{center}
\caption{Theoretical (solid lines) and experimental discharge
current densities for three devices with different average
nanocrystal diameters and nanocrystal densities as a function of
flat-band voltage shift.} \label{dischargedensities}
\end{figure}

The discharge currents have been measured through time decay of
flat-band voltage shift for three devices as shown in Fig.
\ref{dischargedensities}. The measured $\Delta V_{FB}$ is used to
calculate the stored charge using Eq. \ref{hysteresisvoltage} and
divided by the time duration between measurements to calculate the
current. The discharge current is seen to increase with increasing
flat-band voltage shift (or stored charge). Curves for different
nanocrystal sizes show a cross-over behavior. This feature is a
strong evidence for size related discharge of the NC-MOS elements.
Smaller NCs decay faster due to higher quantization energy and
reduced tunnel barrier at low charging ratios of $\Delta
V_{FB}/\Delta V_{max}$. However, if the flat-band voltage shift is
close to $\Delta V_{max}$  or $n_e/N_{NC}\simeq 1$, the
quasi-Fermi level increases rapidly as shown in the inset of Fig.
\ref{DOSNC}. As $\Delta V_{max}$ is proportional to NC density
$N_{nc} \propto 1/d_{nc}^3$, smaller NCs have higher density and
$\Delta V_{max}$. At a given $\Delta V_{FB}$, stored charge per NC
is larger for lower density (large diameter) NCs. Therefore, they
may have a larger quasi-Fermi energy for large NCs than smaller
NCs at a given $\Delta V_{FB}$. Since carriers near the
quasi-Fermi level dominate the discharge, the discharge current
increases rapidly when $\Delta V_{FB}\rightarrow \Delta V_{max}$
as is clearly seen in the data for the device with 7.4 nm diameter
NCs, for which $\Delta V_{max}=1.7$ V. This is in accordance with
the numerical solution shown in solid curves of
Fig.\ref{dischargedensities}. The cross-over behavior shows that
NC size and density as well as total stored charge play an
important role in determination of the charge decay rate.

\begin{figure}[h]
\begin{center}
\resizebox{7 cm}{!}{\includegraphics{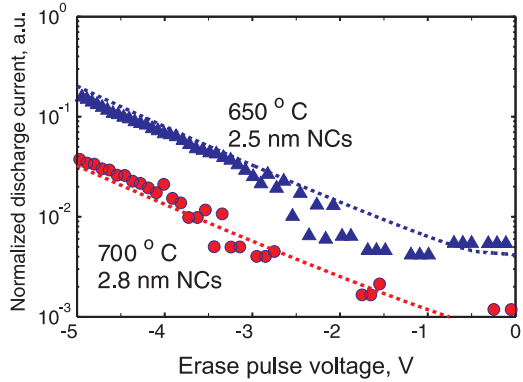}}
\end{center}
\caption{Normalized discharge currents for 2.5 (triangles) and 2.8
nm (circles) diameter NC-MOS devices. Smaller NCs can be erased
faster due to reduced tunnel barrier. Dotted lines are numerical
simulations obtained by only changing the NC size after fitting
the barrier height and width for one of the curves. }
\label{discharge}
\end{figure}

Nanocrystal discharge currents have been measured as a function of
erase pulse voltage. After NC-MOS devices with NC diameters of 2.5
nm and 2.8 nm have been charged by 5 V write pulses to a flat-band
voltage shift of 1.6 V, erase pulses of duration $\tau_{e}=1$ sec
have been applied and flat-band voltage shift has been recorded.
The discharge currents are shown in Fig. \ref{discharge}. The
increased currents for smaller NCs quantitatively confirm the
prediction of numerical calculation. The data of Fig.
\ref{discharge} suggest that the discharge is indeed dominated by
direct tunnelling.

\begin{figure}[h]
\begin{center}
\resizebox{7 cm}{!}{\includegraphics{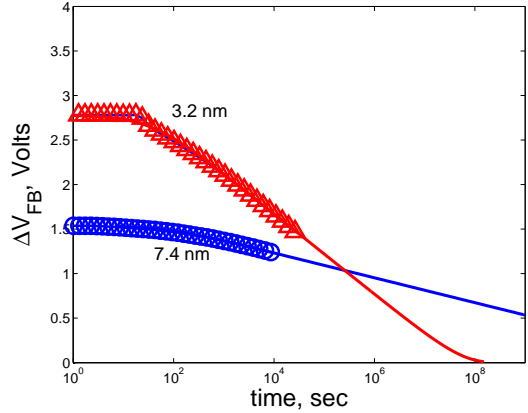}}
\end{center}
\caption{ Smaller NCs with an average diameter of 3.2 nm
(triangles) decay faster than those with an average diameter of
7.4 nm (circles). Solid lines are numerically calculated curves
based on the presented model. } \label{retexample}
\end{figure}

The decay of the charge stored in the NCs has also been recorded
for the NC-MOS capacitors with different NC diameters as a
function of time. The decay of the flat-band voltage shift is
fitted using numerically calculated discharge currents as seen in
Fig. \ref{retexample}. As can be seen, the model predicts the
decay of charge for both short and longer time scales. It is seen
in the time domain also that smaller NCs decay faster than larger
NCs.

\section{Conclusions}
 In conclusion, we have proposed a charge
storage and retention model for nanocrystal MOS memory devices and
compared it with experimental results. The model envisions storage
of carriers in quantized energy levels of NCs. The escape of
carriers is modelled by direct tunnelling out of the NCs to the
substrate. The model can be used to predict the effect of various
design parameters such as NC size and density on retention time.
The model also correctly predicts the super-exponential charge
decay commonly observed in NC memory devices. For NC-MOS
capacitors containing Ge NCs fabricated by the PECVD technique, NC
size related quantum confinement is found to play a role in the
retention of charges. This is an alternative model to surface trap
related carrier storage. The model agrees well with the
experimental results, and gives useful insight to NC-MOS memory
device design.

\begin{acknowledgements}
This work is partially supported by the EU FP6 project SEMINANO
under the contract NMP4 CT2004 505285  and by TUBITAK under
contract No 103T115. Thanks are due to M. Willander of
G\"{o}teburg University for supplying the oxidized silicon wafers.
\end{acknowledgements}

\end{document}